\documentclass[a4paper,11pt]{article}

\usepackage[normalem]{ulem}
\usepackage{color}

\usepackage{pos}

\title{Two-flavored heavy-light mesons' nuclear bound states}

\author*[a]{G.~N.~Zeminiani}
\author[a]{K.~Tsushima}

\affiliation[a]{Laborat\'orio de F\'isica Te\'orica e Computacional (LFTC),
Programa de P\'{o}sgradua\c{c}\~{a}o em Astrof\'{i}sica e F\'{i}sica Computacional,
Universidade Cidade de S\~ao Paulo (UNICID),\\
  01506-000, S\~ao Paulo, SP, Brazil}

\emailAdd{guilherme.zeminiani@gmail.com}
\emailAdd{kazuo.tsushima@gmail.com,
kazuo.tsushima@cruzeirodosul.edu.br}

\abstract{We calculate the
$B^-$-, $\overline{B^0}$-, $D^+$-, $D^0$-, $K^-$-,
and $\overline{K^0}$-$^{12}$C bound state energies by solving the Klein-Gordon (K;G.) equation
in momentum space, and also obtain the corresponding
coordinate space radial wave functions.
The strong Lorentz scalar and vector potentials
in $^{12}$C are calculated using a local density approximation,
where the scalar potentials are obtained based on the mass shifts of
the respective mesons in nuclear matter.
The mesons' mass shifts, the $^{12}$C nuclear density distributions, the strong
potentials as well as the Coulomb potentials, are calculated by
the quark-meson coupling (QMC) model.}

\FullConference{The 21st International Conference on Hadron Spectroscopy and Structure (HADRON2025)\\
 27 - 31 March, 2025\\
Osaka University, Japan\\}

\begin{document}

\hfill {\bf LFTC-25-07/101}

\maketitle

\section{Introduction}

Based on studies of hadron properties under extreme conditions,
one can expect that the Lorentz scalar effective masses
of the mesons would decrease in a nuclear medium, as a consequence of
partial restoration of chiral symmetry~\cite{Krein:2010vp}.
This negative mass shift (Lorentz scalar) of the meson can be regarded as
an attractive Lorentz scalar potential, and when the sum of the Lorentz scalar
and the Lorentz vector potentials (total potential in a nonrelativic picture)
is attractive enough, the mesons can be bound to atomic nuclei.

We study the $B^-$-, $\overline{B^0}$-, $D^+$-, $D^0$-, $K^-$-,
and $\overline{K^0}$-$^{12}$C systems to investigate
the possibility of forming nuclear bound states.
(A similar study was performed for the $D$-mesons~\cite{Tsushima:1998ru}.)
For this purpose, we first solve the Klein-Gordon (K.G.) equation in momentum space
for each orbital angular momentum $\ell$, and then, also
calculate the corresponding coordinate space radial wave functions.

\section{Strong nuclear and Coulomb potentials}

The scalar ($V_\sigma$) and vector ($V_{\omega,\rho}$) potentials, felt
by the mesons in nuclear matter are calculated by the quark-meson coupling
(QMC) model~\cite{Guichon:1987jp,Tsushima:2020gun}
as a function of nuclear matter density.
We use a local density approximation to obtain the meson-nucleus potentials
in $^{12}$C as a function of the distance $r$ from the center of the nucleus,
with the nuclear density distribution in $^{12}$C calculated by the QMC
model~\cite{Saito:1996sf}. The Coulomb potentials in $^{12}$C, when the mesons
are absent, are also self-consistently calculated by the QMC model~\cite{Saito:1996sf}.

\begin{figure}[htb]
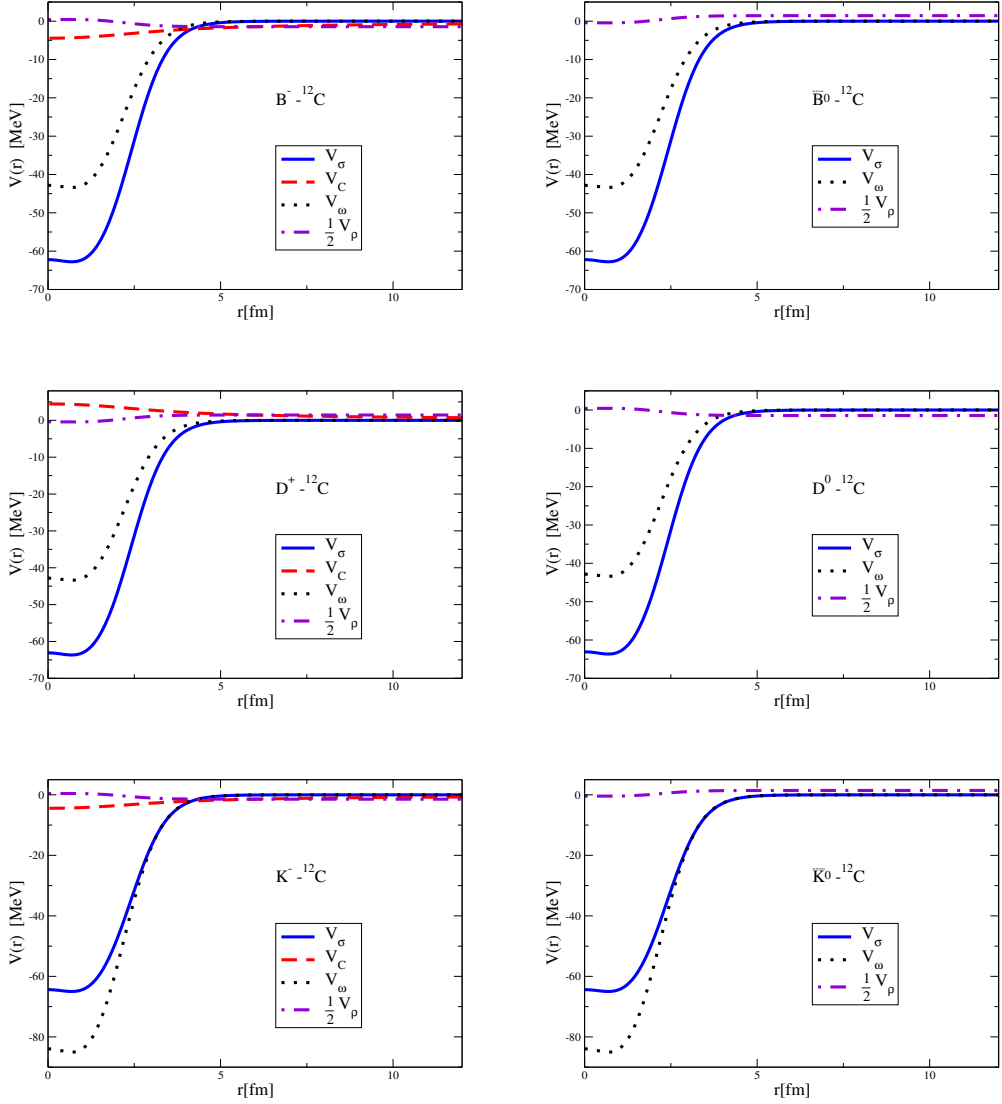
%
\centering
\includegraphics[width=6cm]{B-_C12_pots.eps}
\hspace{5ex}
\includegraphics[width=6cm]{Bbar0_C12_pots.eps}
\\
\vspace{5ex}
\centering
\includegraphics[width=6cm]{D+_C12_pots.eps}
\hspace{5ex}
\includegraphics[width=6cm]{D0_C12_pots.eps}
\\
\vspace{5ex}
\centering
\includegraphics[width=6cm]{K-_C12_pots.eps}
\hspace{5ex}
\includegraphics[width=6cm]{Kbar0_C12_pots.eps}
 \caption{Nuclear and Coulomb potentials for the $B^-$-$^{12}$C (top left),
 $\overline{B^0}$-$^{12}$C (top right), $D^+$-$^{12}$C (center left), $D^0$-$^{12}$C (center right), $K^-$-$^{12}$C (bottom left) and $\overline{K^0}$-$^{12}$C (bottom right) systems.}
\label{nuclpots}
\end{figure}

When the repulsion from the vector potentials is sufficiently large,
the meson cannot form a bound state with the nucleus.
In Fig.~\ref{nuclpots} we present the nuclear and Coulomb potentials in $^{12}$C
felt by $B^-$-, $\overline{B^0}$-, $D^+$-, $D^0$-, $K^-$-,
and $\overline{K^0}$ mesons,  namely, the scalar $V_\sigma$, vector
$V_{\omega,\rho}$ and Coulomb $V_C$ potentials.
In this study we neglect the meson widths, that would enter the calculations
as the imaginary part of the meson-nucleus potentials.

\section{Nuclear bound states}

\begin{figure}[htb!]
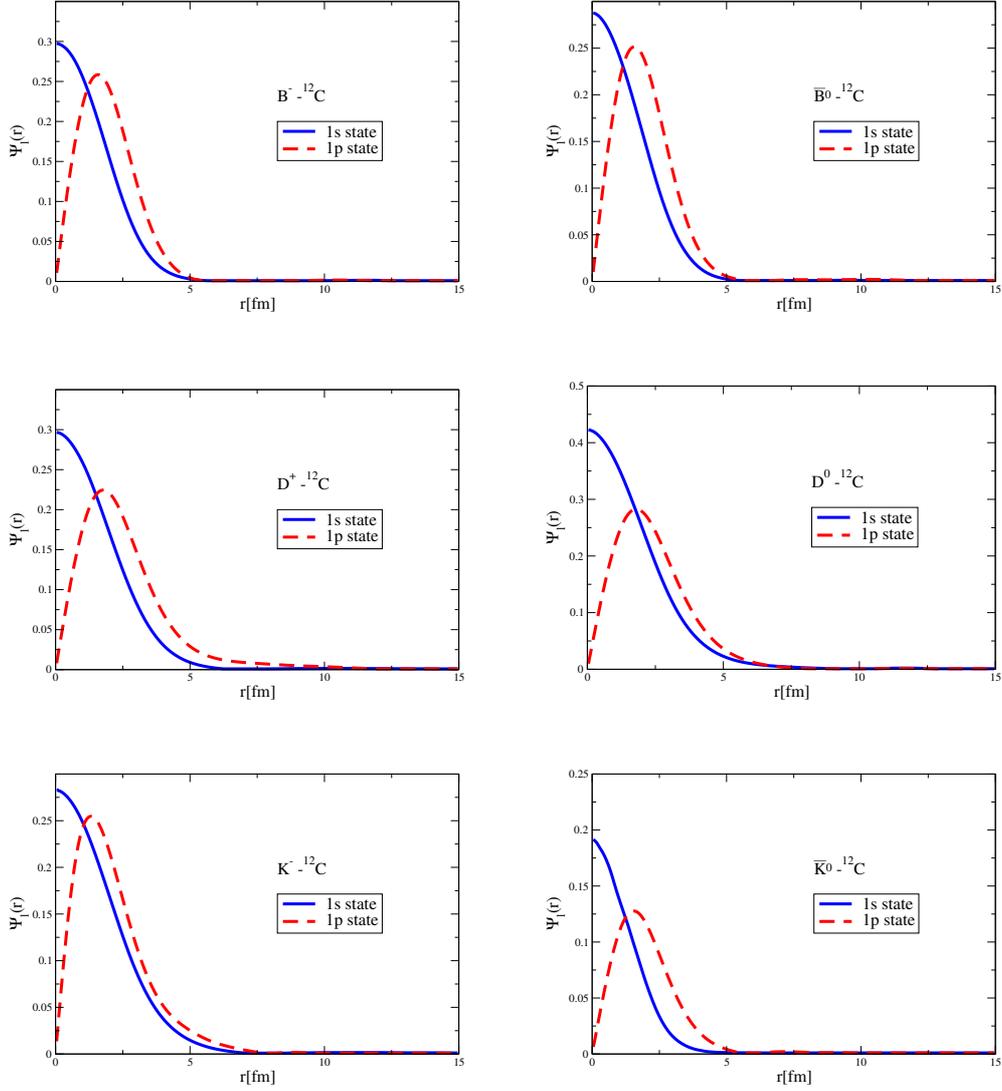
%
\centering
\includegraphics[width=6cm]{Psir_B-_C12_1s1p.eps}
\hspace{5ex}
\includegraphics[width=6cm]{Psir_Bbar0_C12_1s1p.eps}
\\
\vspace{5ex}
\centering
\includegraphics[width=6cm]{Psir_D+_C12_1s1p.eps}
\hspace{5ex}
\includegraphics[width=6cm]{Psir_D0_C12_1s1p.eps}
\\
\vspace{5ex}
\centering
\includegraphics[width=6cm]{Psir_K-_C12_1s1p.eps}
\hspace{5ex}
\includegraphics[width=6cm]{Psir_Kbar0_C12_1s1p.eps}
 \caption{Coordinate-space 1s- and 1p-state wave functions for the $B^-$-$^{12}$C (top left),
 $\overline{B^0}$-$^{12}$C (top right), $D^+$-$^{12}$C (center left), $D^0$-$^{12}$C (center right), $K^-$-$^{12}$C (bottom left) and $\overline{K^0}$-$^{12}$C (bottom right) systems.}
\label{wvf}
\end{figure}

We numerically solve the K.G. equation in momentum space for each value of the angular
momentum $\ell$ of the meson-nucleus systems to calculate the bound-state energies
of each energy level.
The partial wave solutions for the bound-state energies and the corresponding
momentum-space wave functions of the K.G equation require the decomposition of
the nuclear and Coulomb potentials into the angular momentum $\ell$ waves.
We apply the double spherical Bessel transform on the coordinate-space potentials
to obtain the $\ell$-wave decomposition of the potentials in momentum space.
The binding energies $E_{n\ell}$ of
the $B^-$-$^{12}$C, $\overline{B^0}$-$^{12}$C,
$D^+$-$^{12}$C, $D^0$-$^{12}$C, $K^-$-$^{12}$C and $\overline{K^0}$-$^{12}$C systems
are presented in Table~\ref{tblbs}.
In addition, the corresponding coordinate-space wave functions are
presented in Fig.~\ref{wvf}, which are obtained by a spherical Bessel
transform from the normalized wave functions in momentum space.

\begin{table}[htb!]
\caption{\label{tblbs} $B^-$-, $\overline{B^0}$-, $D^+$-, $D^0$-, $K^-$-,
and $\overline{K^0}$-$^{12}$C bound state energies in MeV. }
\begin{center}
\begin{tabular}{ll|r|r}
  \hline \hline
  & & \multicolumn{2}{c}{$E_{n\ell}$ (MeV)} \\
\hline
& $n\ell$ & $B^-$ & $\overline{B^0}$ \\
\hline
$^{12}_{B}\text{C}$ 
& 1s & -66.47 & -65.99 \\
& 1p & -42.69 & -39.94 \\
\hline
& $n\ell$ & $D^+$ & $D^0$ \\
\hline
$^{12}_{D}\text{C}$
& 1s & -38.57 & -42.62 \\
& 1p & -32.31 & -36.8 \\
\hline
& $n\ell$ & $K^-$ & $\overline{K^0}$ \\
\hline
$^{12}_{K}\text{C}$ 
& 1s & -33.24 & -32.9 \\
& 1p & -26.87 & -25.24 \\
\hline
\end{tabular}
\end{center}
\end{table}

\section{Summary and Conclusion}

We have presented the results calculated for the bound-state energies of
the $B^-$-, $\overline{B^0}$-, $D^+$-, $D^0$-, $K^-$-,
and $\overline{K^0}$-$^{12}$C systems
by solving the Klein-Gordon equation in momentum space.
The nuclear and Coulomb potentials were calculated
by the quark-meson coupling (QMC) model.
The present results were obtained neglecting the imaginary parts of the meson-nucleus
potentials. Under this restriction, they predict that the ($B^-,\overline{B^0}$), ($D^+,D^0$)
and ($K^-,\overline{K^0}$) mesons should form bound states with the $^{12}$C nucleus.
In the near future, we will extended the present study to include
the imaginary parts of the meson-nucleus potentials,
and also treat the other nuclei.

\newpage

\begin{acknowledgments}
G.N.Z.~was supported by the Coordena\c{c}\~ao de Aperfei\c{c}oamento de Pessoal
de N\'ivel Superior-Brazil (CAPES), and FAPESP Process No.~2023/07313-6.
K.T. was
supported by Conselho Nacional de Desenvolvimento
Cient\'{i}ıfico e Tecnol\'{o}gico (CNPq, Brazil), Processes
No. 304199/2022-2, and~FAPESP No. 2023/07313-6.
This work was also supported by, Instituto Nacional de
Ci\^{e}ncia e Tecnologia - Nuclear Physics and Applications
(INCT-FNA), Brazil, Process No. 464898/2014-5.
\end{acknowledgments}


\begin{thebibliography}{99}

\bibitem{Krein:2010vp}
G.~Krein, A.~W.~Thomas and K.~Tsushima,
Phys. Lett. B \textbf{697}, 136 (2011).

\bibitem{Tsushima:1998ru}
K.~Tsushima {\it et. al.,} 
Phys. Rev. C \textbf{59}, 2824 (1999).

\bibitem{Cobos-Martinez:2022fmt}
J.~J.~Cobos-Mart\'\i{}nez, G.~N.~Zeminiani and K.~Tsushima,
Phys. Rev. C \textbf{105}, 025204 (2022).
  
\bibitem{Guichon:1987jp}
P.~A.~M.~Guichon,
Phys. Lett. B \textbf{200}, 235 (1988).
  
\bibitem{Tsushima:2020gun}
K.~Tsushima,
PTEP \textbf{2022}, 043D02 (2022).

\bibitem{Saito:1996sf}
K.~Saito, K.~Tsushima and A.~W.~Thomas,
Nucl. Phys. A \textbf{609}, 339 (1996).

\end{thebibliography}
\end{document}